\documentclass[11pt]{article}
\fontfamily{times}
\usepackage{graphicx}
\usepackage{geometry}

\usepackage{natbib}

\newcommand{\mnras}[1]{Monthly Notices of the Royal Astronomical Society}
\newcommand{\apj}[1]{Astrophysical Journal}
\newcommand{\apjl}[1]{ApJL}
\newcommand{\apjs}[1]{ApJS}
\newcommand{\aj}[1]{AJ}
\newcommand{\aap}[1]{Astronomy \& Astrophysic}
\newcommand{\aaps}[1]{Astronomy and Astrophysics Supplement Series}
\newcommand{\araa}[1]{Ann. Rev. A\&A}
\usepackage[font=footnotesize,labelfont=bf]{caption}
\newcommand{\templatefigures}[1]
{\noindent
\begin{minipage}{2cm}
\begin{center}
  \centering
	\vspace{-1cm}
  \includegraphics[scale=0.2]{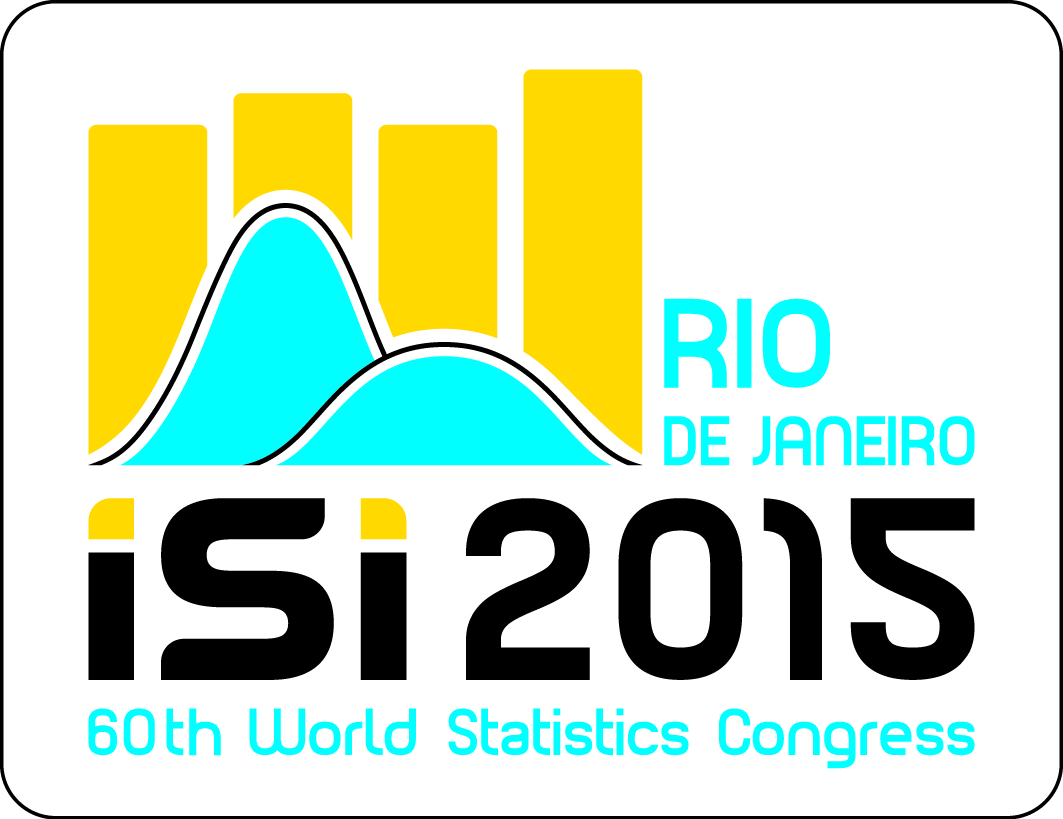}\\
\end{center}
\end{minipage}
\quad
\begin{minipage}{12cm}
\hspace*{6.8cm}
\end{minipage}
\quad
\begin{minipage}{2cm}
\begin{center}
\vspace{-0.9cm}
\includegraphics[scale=0.35]{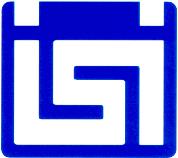}\\
\end{center}

\end{minipage}

\vskip0.2cm
}

\begin{document}
\templatefigures{}

\small{

\begin{center}
%The title should be centred and in bold letters. It should be informative but not too long (preferably no more than two lines).
\textbf{Clustering with phylogenetic tools in astrophysics}
\end{center}

\begin{center}
{Didier Fraix-Burnet*}\\
{Univ. Grenoble Alpes / CNRS, IPAG, Grenoble, France - didier.fraix-burnet@univ-grenoble-alpes.fr}\\ 
% \vspace{0.5cm}
% 
% {Name Surname}\\
% {Institution, City, Country - e-mail address}\\

\end{center}

\begin{center}
{\bf Abstract}
\end{center}

\setlength{\parindent}{0pt}

Phylogenetic approaches are finding more and more applications outside the field of biology. Astrophysics is no exception since  an overwhelming amount of multivariate data has appeared in the last twenty years or so. In particular, the diversification of galaxies throughout the evolution of the Universe quite naturally invokes phylogenetic approaches. We have demonstrated that Maximum Parsimony brings useful astrophysical results, and we now proceed toward the analyses of large datasets for galaxies. In this talk I present how we solve the major difficulties for this goal: the choice of the parameters, their discretization, and the analysis of a high number of objects with an unsupervised NP-hard classification technique like cladistics.
\\

}\\

\setlength{\parindent}{0pt}

% figures and tables should be centered

{\bf 1. Introduction}

How do the galaxy form, and when? How did the galaxy evolve and transform themselves to create the
diversity we observe? What are the progenitors to present-day galaxies? To answer these big questions,
observations throughout the Universe and the physical modelisation are obvious tools. But between these,
there is a key process, without which it would be impossible to extract some digestible information from the
complexity of these systems. This is classification. 

One century ago, galaxies were discovered by Hubble. From images obtained in the visible range of
wavelengths,   he   synthetised  his  observations  through  the  usual   process:   classification.   With   only  one
parameter (the shape) that is qualitative and determined with the eye, he found four categories: ellipticals,
spirals, barred spirals and irregulars. This is the famous Hubble classification. He later hypothetized relationships between these classes, building the Hubble Tuning Fork.

The Hubble classification has been refined, notably by de Vaucouleurs, and is still used as the only global
classification of galaxies. Even though the physical relationships proposed by Hubble are not retained any
more, the Hubble Tuning Fork is nearly always used to represent the classification of the galaxy diversity
under its new name “the Hubble sequence” \citep[e.g.][]{DelgadoSerrano2010}. Its success is impressive and can be
understood by its simplicity, even its beauty, and by the many correlations found between the morphology of
galaxies   and   their  other  properties.  And  one  must   admit   that  there  is   no
alternative up to now, even though both the Hubble classification and diagram have been recognised to be
unsatisfactory. Among the most  obvious flaws of this classification,  one must  mention its monovariate, qualitative,
subjective and old-fashioned nature, as well as the difficulty to characterise the morphology of distant
galaxies.

The first two most significant multivariate studies
were by  \citet{Watanabe1985}  and  \citet{Whitmore1984}. Since the year 2005, the number of studies
attempting to go beyond the Hubble classification has increased largely. Why, despite of this, the Hubble
classification and its sequence are still alive and no alternative have yet emerged \citep{Sandage2005}?

My feeling is that the results of the multivariate analyses are not easily integrated into a one-century old practice of modeling the observations. In addition, extragalactic objects like galaxies, stellar clusters or stars do evolve. Astronomy now provides data on very distant objects, raising the question of the relationships between those and our present day nearby galaxies. Clearly, this is a phylogenetic problem.

Astrocladistics\footnote{http://astrocladistics.org} aims at exploring the use of phylogenetic tools in astrophysics \citep{jc1,jc2}. We have proved that Maximum Parsimony (or cladistics) can be applied in astrophysics and provides a new exploration tool of the data \citep{FDC09,Fraix2012,Cardone2013}. As far as the classification of galaxies is concerned, a larger number of objects must now be analysed. In this paper, I detail how we tackle the technical difficulties.
\\

{\bf 2. Parameters as tracers of the evolutionary history: an illustration with stellar tracks }

%Comparison between partitioning and phylogenetic. Mettre peut-être aussi une mention de l'article de 2012 ou on trouve des accords corrects.\\

\begin{figure}[h]
\begin{center}
  \includegraphics[width=0.25\linewidth]{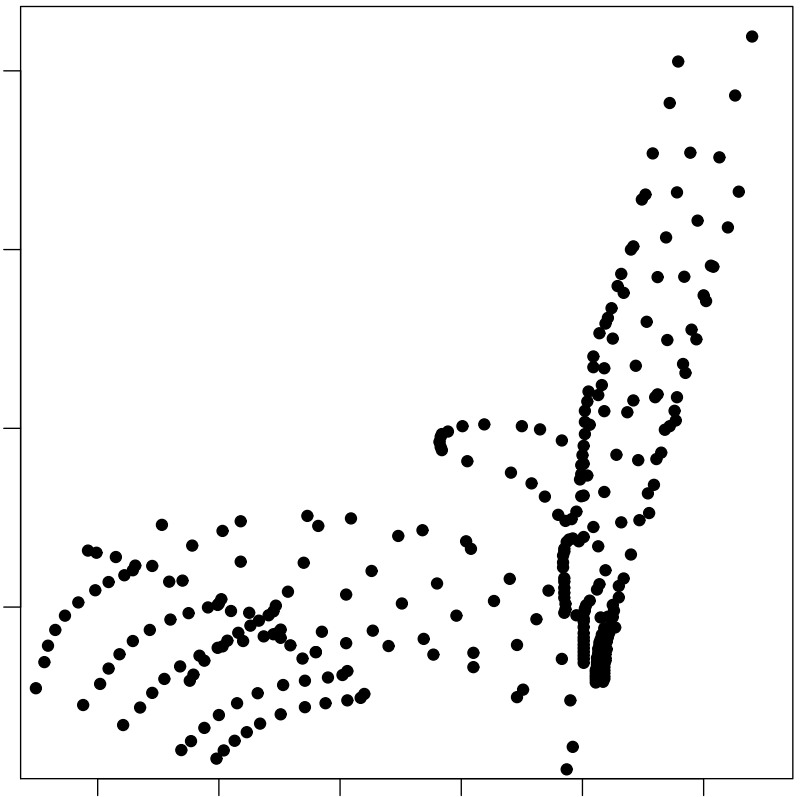}
  \includegraphics[width=0.28\linewidth]{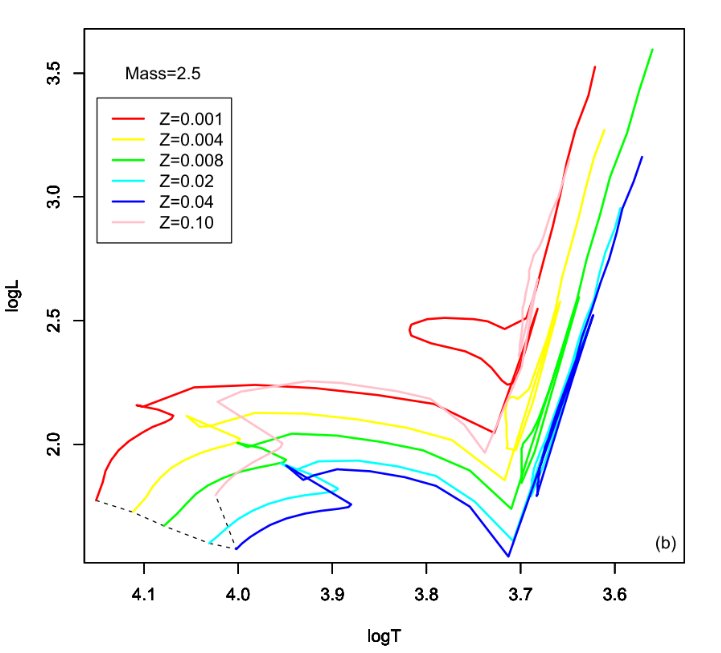}
\caption{Left: a sample data points. Right: the full stellar evolutionary tracks corresponding to the data points (51 points per track). Each track corresponds to the evolution of a star with an initial metallicity Z given in the inlet. All stars in this sample have an initial mass of 2.5 times the solar mass. The dotted line on the bottom left is the Main Sequence where the computations start.}
\label{Fig:HRdata}
\end{center}
\end{figure}

Let us consider the data shown on the left plot in Fig.~\ref{Fig:HRdata}. How many groups are there? 

The answer is given on the right plot of  Fig.~\ref{Fig:HRdata}: six. These plots give the luminosity $\log L$ as a function of the surface temperature of the star $\log T_{eff}$. They are known as the Hertzsprung-Russell (HR) diagram and are made using the Geneva stellar evolutionary models \citep{grid2,grid5,grid3, grid4, grid1}
%(Charbonnel et al 1993; Mowlavi et al 1998; Schaerer et al 1993a, b; Schaller et al 1992) 
that compute the stellar parameters for stars of masses from M = $0.8$ to $120$ Mo and metallicities Z from $= 0.001$ to $0.1$ at different ages (time steps or evolutionary stages). We have chosen six values of Z and one mass (see Fig.~\ref{Fig:HRdata}). Each of our six tracks has 51 time steps which are our data points. The simulations provide 11 additional parameters which give the chemical composition of the stars. The analyses below are made with the 13 parameters.

\begin{figure}[h]
\begin{center}
  \includegraphics[width=0.25\linewidth]{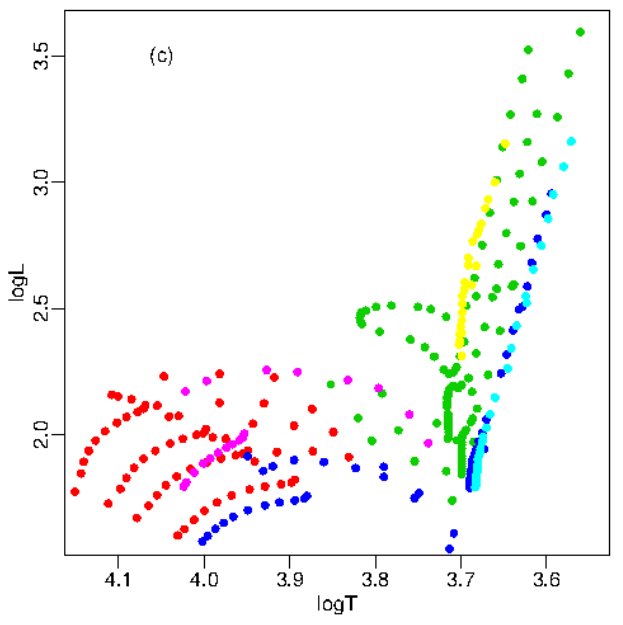}
  \includegraphics[width=0.28\linewidth]{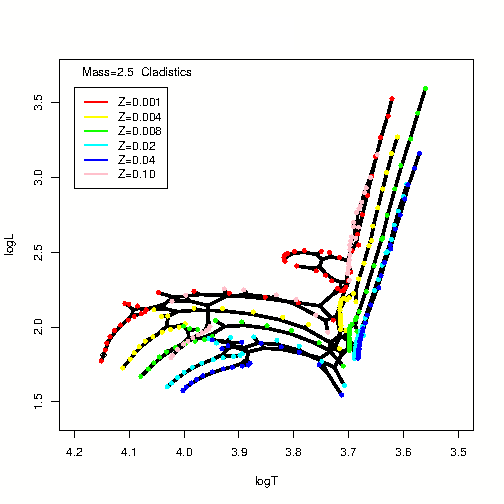} 
\end{center}
\caption{Left: the result of a k-medoids analysis assuming six clusters. The colors used to distinguish the groups are arbitrary. Right: the result of the cladistics analysis. The colors of the points corresponds to the true families of stars characterized by their initial metallicity (inlet) . The black lines are the branches of the tree indicating the relationships between the data points. In other words, this plot is the projection of the tree on the bivariate diagram $\log L$ vs $\log T_{eff}$.}
\label{Fig:HRresults}
\end{figure}

A clustering approach like k-medoids would yield the result shown on the left of Fig.~\ref{Fig:HRresults}, while the cladistic analysis yields the result shown on the right plot. Obviously, cladistics does a pretty good job at recovering the six evolutionary tracks. But there are some discrepancies which are due to the behaviour of the parameters with respect to the evolution. It can be seen below in Fig.~\ref{Fig:HRparam} that a lot of parameters (shown here only for two tracks) are constant during the first half of the evolution, and at least one parameter shows a reversal (circle). 

\begin{figure}[h]
\begin{center}
 \includegraphics[width=0.5\linewidth]{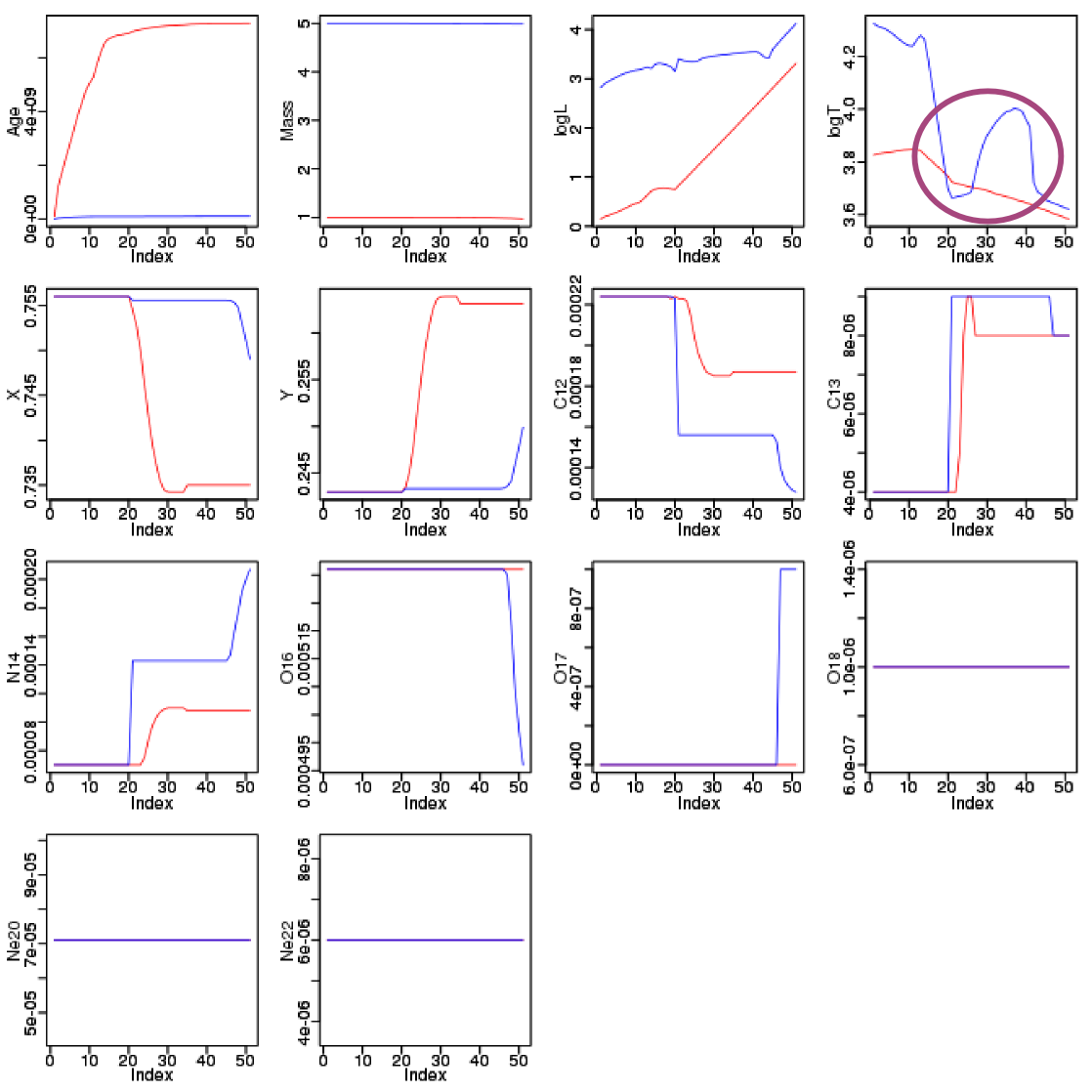}
\caption{Evolution of the 13 parameters for two tracks (i.e. two metallicities) as a function of the time steps. The circle indicates a reversal in the evolution of the parameter $\log T_{eff}$.}
\label{Fig:HRparam}
\end{center}   
\end{figure}

Reversals, convergences or parallel evolutions are called homoplasies and tend to destroy the phylogenetic signal. The Maximum Parsimony technique minimizes these homoplasies and maximizes the synapomorphies.

So the crucial point in phylogenetic methods is the selection of the parameters that are synapomorphies, which keep the traces of the ancestorships. There is no method to detect them except trial-and-error approaches, and some a priori knowledge help to eliminate obvious homoplasies.
\\

{\bf 3. Discretizing the Parameters}

Cladistics has initially been devised for quatlitative and discrete parameters (called characters), each bin representing an evolutionary state for that parameter. The use of continuous data has long been debated but seems now to be accepted \citep{Goloboff2006} and even justified \citep{TF09}. 

Since there is yet no phylogenetic algorithm using continuous parameters and adapted to the kind of evolution processes for galaxies, we have to bin our data. Of course, there is no unique choice:
\begin{enumerate}
   \item taking a small number of bins could reduce the number of distinct objects, but we have noted that the result depends significantly on the precise number of bins below, say, fifteen bins. In any case, should the bin be equals~? And if not, where should we place the limits?
  \item taking more bins more or less respects the continuous nature of the parameters, but this may question the meaning of the bins as evolutionary states. 
\end{enumerate}

We have always adopted the second choice, and used most often 32 bins, noting that in any case this is the physical interpretation that ensures the relevance of the binning. 
\\

{\bf 4. Unkown Species and the High Number of Individuals: the Example of Galaxies}

Maximum Parsimony is a NP-hard problem: it looks for all the possible arrangements of objects on a tree and then selects the most parsimonious one. Above a thousand individuals, this takes too much CPU time, especially if we have to repeat the analysis a large number of times, either to select the most satisfactory subsets of parameters, or to assess the robustness of the result via a bootstrap method.

The problem here is that there is no multivariate classification of galaxies so that the species are not known at all. We thus have to make the hypothesis that each individual is an exemplar of a class to be identified. 

\begin{figure}[h]
\begin{center}
    \includegraphics[width=0.5\linewidth]{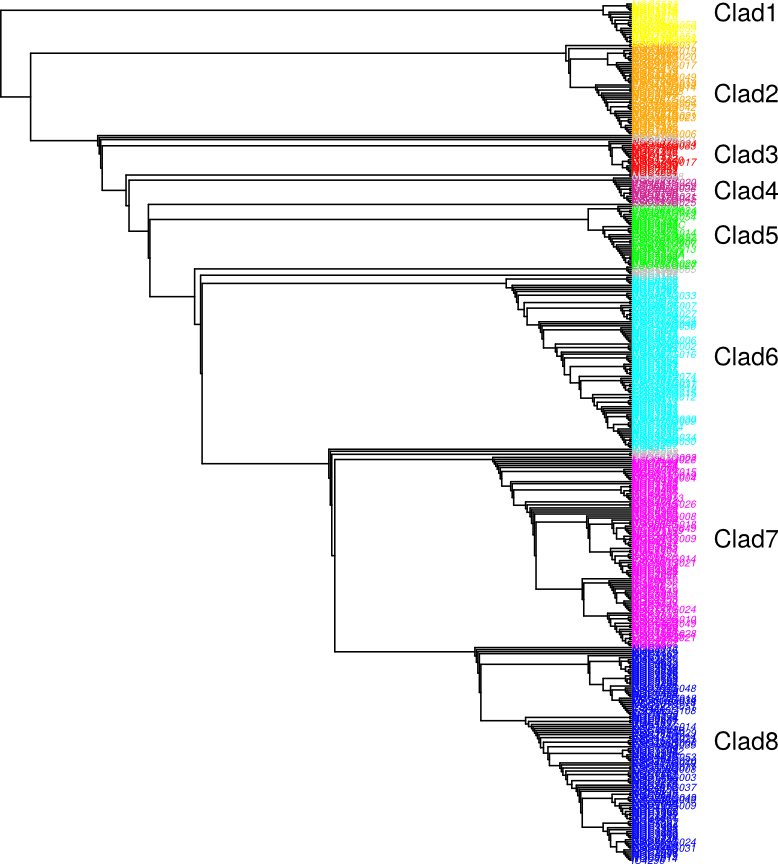} 
\hfill
 \includegraphics[width=0.3\linewidth]{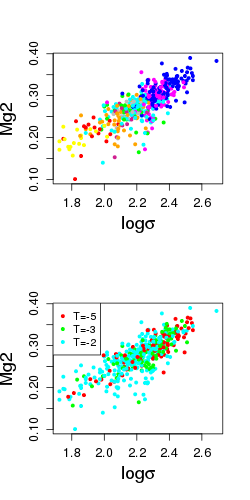}
\caption{Left: tree of galaxies, from \citet{Fraix2012}. Top right: metallicity indicator Mg$_2$ vs the velocity dispersion $\log\sigma$ with the colors corresponding to the groups of the tree to the left. Bottom right: same diagram with the color-coded morphological classes.}
\label{Fig:galaxies}
\end{center}
\end{figure}

Up to now, we have analysed only a few hundreds galaxies at a time, as shown on the tree below \citep[Fig.~\ref{Fig:galaxies} left; ][]{Fraix2012}. The groups or classes are defined from the tree, and the average properties for each class are computed. This constitutes a first and partial phylogeny of galaxies that must be extended.

One important consequence of our analysis is that the correlations (scaling relations) must be revisited since they often vary depending on the class or even disappear within the classes. A spectacular example is the well-known scaling relation between the velocity dispersion $\sigma$ and the metallicity (here the Mg$_2$ index). The Maximum Parsimony result (top right in Fig.~\ref{Fig:galaxies}) shows that the correlation is absent in most if not all of the groups. Indeed, the global correlation appears to be caused by the diversification of galaxies: the relative position of the groups in the plot follows exactly the relative position on the tree. This is a clear evidence that a confounding factor explains this correlation. 

On the contrary, the three morphological classes present in the sample show the same correlation (bottom right diagram in Fig.~\ref{Fig:galaxies}). This would indicate that all galaxies have the same history, which is not plausible. 
\\

{\bf 4. Selecting Smaller Representative Subsets}

Our current goal is to extend the application of Maximum Parsimony to much larger samples of galaxies. To keep the problem computationally tractable, there are two pathways:
\begin{enumerate}
   \item keep the already established phylogeny, define an exemplar for each class, and add new objects progressively;
   \item define small sub-samples, determine the phylogenies and combine them with supertree techniques.
\end{enumerate}

The first approach is easier with discrete data since a species can be characterized precisely. For continuous data, the class limits are necessarily overlapping, and a parameter specific to a class is defined by its quantiles. This requires a sufficient number of individuals for these quantiles to be representative enough of a given class. Hence this approach is quite interesting when a significant fraction of the whole sample has been analysed, the already established phylogeny serving as a backbone tree to contrain the Maximum Parsimony search.

The second approach randomly selects a tractable subset of the data. For this, it appears necessary that the subset is representative of the whole sample, ideally with the same distribution for each parameter. Subsequently, a supertree can be built using all the sub-trees if the latter are sufficienctly robust.

The two approaches are indeed complementary and can be run in parallel.
\\

{\bf 5. Conclusion}

To be able to replace the one-century old Hubble Tuning Fork with a multivariate phylogenetic classification, the phylogenetic approach has now proven to be a viable pathway. To tackle larger samples of galaxies, astrocladistics faces several challenges which seem to be surmountable.

% \begin{enumerate}
%    \item the selection of the parameters that should be synapomorphies;
%    \item the necessary binning of the parameter values for the softwares: how many bins should we consider? Should they be equal?
%    \item the unknown species, requiring the assumption that each individuals are exemplars of classes to be found (unsupervised classification)
%    \item the number of individuals is too high for cladistics which is a NP-hard problem.
% \end{enumerate}

The Maximum Parsimony is not the only phylogenetic tool. More generally, distance-based techniques -- phylogenetic, partitioning or hierarchical -- are much more computationally efficient. However, I think that the cladistics method allows for a clearer conceptual clarification of the diversification processes. This was obvioulsy a necessary step before performing large scale multivariate clustering analyses of complex objects in evolution. 
\\

%{\bf References}\\

%   \bibliographystyle{bibformatDFB}
% \bibliography{/home/fraix/Documents/biblio/JabRefDatabase/JabRef,/home/fraix/Documents/biblio/JabRefDatabase/publicationsDFB,/home/fraix/Documents/articles/livre_astrocladistique/version3/livrebiblio,/home/fraix/Documents/biblio/JabRefDatabase/fundplanebiblio,/home/fraix/Documents/articles/etoilesHR/etoilesHR}
% %\bibliography{/home/fraix/Documents/biblio/JabRefDatabase/JabRef,/home/fraix/Documents/biblio/JabRefDatabase/publicationsDFB,/home/fraix/Documents/articles/etoilesHR/etoilesHR.bib}
{\small%footnotesize

}

\end{document}